\documentstyle[epsfig,11pt]{article}
\newcommand{\be}{\begin{equation}}
\newcommand{\ee}{\end{equation}}
\newcommand{\ben}{\begin{eqnarray}}
\newcommand{\een}{\end{eqnarray}}
\newcommand{\bc}{\begin{center}}
\newcommand{\ec}{\end{center}}

\def\rmo{{\rm o}}

\def\ltsima{$\; \buildrel < \over \sim \;$}
\def\simlt{\lower.45ex\hbox{\ltsima}}
\def\gtsima{$\; \buildrel > \over \sim \;$}
\def\simgt{\lower.45ex\hbox{\gtsima}}

\begin{document}

\title{The relativistic Iron K$\alpha$ line from an
accretion disc onto a static non-baryonic compact object}
\date{\today}
\author{Youjun Lu$^a$\footnote{Current Address: Astrophysical Sciences
Department, Princeton University, NJ 08544, USA} ~and Diego F. Torres$^b$\\
{\small $^a$Center for Astrophysics, University of
Sci. \& Technology of China, Hefei, Anhui 230026, P. R. China}\\
{\small $^b$Physics Department, Princeton University, NJ 08544,
USA}}

\maketitle

\begin{abstract}
This paper continues the study of the properties of an accretion
disc rotating around a non-baryonic (assumed super-massive)
compact object. This kind of objects, generically known as boson
stars, were earlier proposed as a possible alternative scenario to
the existence of super-masive black holes in the center of every
galaxy. A dilute boson star has also been proposed as a large part
of the non-baryonic dark matter, flattening galactic rotational
velocities curves. In this contribution, we compute the profile of
the emission lines of Iron; its shape has been for long known as a
useful diagnosis of the space-time geometry. We compare with the
case of a Schwarzschild black hole, concluding that the
differences are observationally distinguishable. \\

PACS Number(s): 04.40.Dg, 98.62.Mw, 04.70.-s
\end{abstract}

\section{Introduction}\label{sec:intro}

It has long been believed that accretion onto massive black holes
is the energy source of active galactic nuclei (AGN). Nearby
galaxies, as the remnants of AGNs, should then harbor massive
black holes in their centers \cite{rees}. Recent observations have
indeed revealed the existence of massive dark objects, believed to
be those massive black holes, in most, if not all, nearby galactic
nuclei \cite{kr95,mag98}. However, those observations only reach a
physical radius 10000 times larger than that of the event horizon
of the putative black hole. As pointed out in Ref.
\cite{Torres:2000dw}, from a dynamical point of view, a single
massive boson star is also compatible with all measurements of
star kinematics around the massive dark object in the most studied
center, that of Milky Way, without requiring the massive dark
object to be a black hole. A proof of the existence of massive
black holes requires unambiguous evidence of the event horizon. It
is then important to address the issue of the emissivity
properties of matter rotating around different central objects.
Ref. \cite{Torres:2002td} made a first step in this sense; to that
paper we refer the reader for further discussions and a more
complete list of references on boson star physics and related
problems. The emissivity properties are much more important than
the kinematic studies, since they can provide a more direct way to
actually
distinguish between different central objects \cite{neutrino}. \\

X-ray studies have revealed the most prominent spectral feature
coming from the central regions of some active galaxies, an
enormous Doppler and gravitational-shifted broad Fe K$\alpha$
line, which, for the first time, showed the existence of a strong
gravity regime \cite{tan95,fab95}. This has been taken as the
strongest evidence for the existence of super-massive black holes,
or even as a sensitive diagnosis of the black hole spin. But, how
would the profile of the emission lines be for those produced in
discs rotating around putative boson stars? Would them present
sufficiently different shapes as to make them distinguishable in
forthcoming observations? These are the questions we address in
this paper.

\section{Photon propagation in the metric of a boson star}\label{sec:photon}

\subsection{Space-time}

The space-time generated by a stationary and spherically symmetric
object can be described by the metric: \be\label{eq:sys1}
ds^2=B(r)dt^2-A(r)dr^2-r^2(d\theta^2+\sin^2\theta d\phi^2), \ee
where $A(r)$ can be written as $1/(1-2GM(r)/rc^2)$, $c$ is the
speed of light, $G$ is the gravitation constant and $M(r)$ is the
mass enclosed within a radius $r$. The simplest solution for
Eq.~(\ref{eq:sys1}) is that representing a Schwarzschild black hole,
for which $B(r)=1-2GM/rc^2$, and $M(r)$ is a constant equal
to the mass of the black hole.\\

Other stellar-like systems described by Eq.~(\ref{eq:sys1}) are
those known as boson stars.  Details of boson star solutions can
be found in the reviews quoted in Ref. \cite{reviews}. For a boson
star, $A(r)$ and $B(r)$ can be calculated only numerically (see
\cite{Torres:2000dw} for explanations, the program herein used is
an adaptation of that mentioned in Refs. \cite{prog}).  \\

The total mass, the radius and all other physical parameters of
the boson star are scaled by the mass of the constituent boson,
$m$. For numerical purposes, then, we can work with dimensionless
units. The radius in a dimensionless form will be $x=m r$, and the
mass, as well, will be transformed into a dimensionless quantity,
$M=M(x) M^2_{\rm Planck}/m$, where $M_{\rm Planck}=\sqrt{{\hbar
c}/{G}}$, $\hbar$ is the Planck constant. This scheme was already
proposed in the first papers on boson stars \cite{RB,KAUP,col86}.
For example, if the massive dark object with mass $2.6\times10^{6}
{\rm M_{\odot} }$ in the Galactic Center is a boson star, then the
constituent boson mass is $m=2.81 \times 10^{-26}$ GeV. The
distance from the center in unit of parsec is $r[{\rm
pc}]=x/m[{\rm GeV}]~6.38 \times 10^{-33}$, the mass in millions of
solar mass is $M[10^6]=M(x)/m[\rm GeV]~1.33 \times 10^{-25}$, and
the gravitational radius $r_{\rm g}=GM/c^2$ is written as $x_{\rm
g}=0.53256$ in dimensionless units. The boson star structure is
numerically computed to obtain $M(x)$ and $B(x)$
\cite{Torres:2000dw}. Two notes are, however, in order. Firstly,
these numbers rely in that we are using a non self-interacting
boson star, for simplicity and because the behavior is generic,
but should be consistently changed if this is not the case
\cite{Torres:2000dw}. By going towards a self-interacting model,
the mass of the constituent boson needed to produce the same
overall mass increases. Secondly, the use of the mass of the
galactic center region is done just to provide a numerical
example, we are not claiming the possible existence of a boson
star in the center of the Galaxy, since apparently the emissivity
properties would conflict with observations (see Ref.
\cite{Torres:2002td}), or at least we know that
that is the case for the simplest accretion disc models.\\

\begin{figure}[t]
\centering
\epsfxsize=0.5\textwidth \epsfysize=0.35\textheight
\epsffile{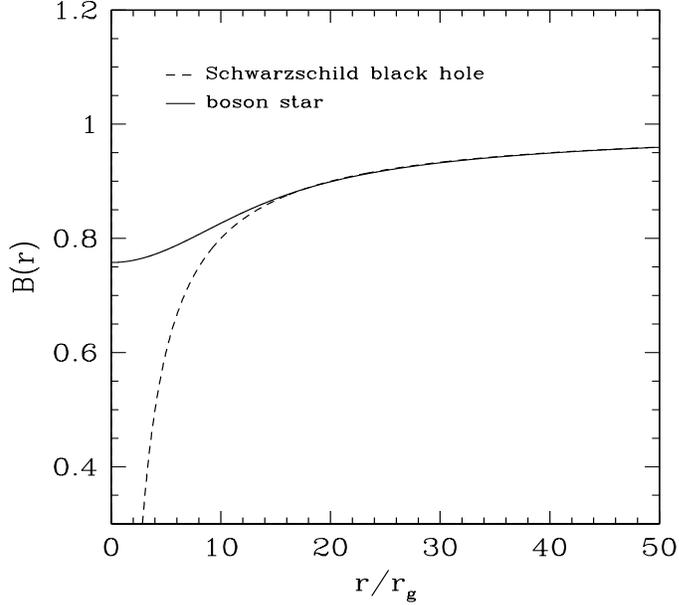} \caption{The metric parameter $B(r)$ as a
function a radius (in unit of the gravitation radius $r_{\rm
g}=GM/c^2$). The solid line represents a boson star and the dash
line represents a Schwarzschild black hole of equal mass. The main
difference appears in the inner region, with $r<15r_{\rm g}$.}
\label{fig:gttfig}
\end{figure}

To conveniently compare the metric generated by a non-rotating
boson star with the Schwarzschild black hole metric, we re-scale
the total mass of the boson star to be $M=1$. We do so by just
dividing $x/0.53256$ and replacing $M(x)$ by $M(x)/0.53256$.
Figure~\ref{fig:gttfig} shows $B(r)$ as a function of $r$ for both
boson star and Schwarzschild black hole. $B(r)$ for both cases are
almost exactly the same when $r>15r_g$, while it slowly decreases
to a constant at the center of a boson star rather than rapidly
decrease to zero at the horizon ($r=2r_{\rm g}$) of a
Schwarzschild black hole. To a distant observer, thus, a photon
emitted from the inner region of a disc extending within a boson
star ($r<15r_g$) suffers much less gravitational redshift than the
one emitted from the disc surrounding a
Schwarzschild black hole.\\

\begin{figure}[t]
\centering \epsfxsize=0.5\textwidth \epsfysize=0.5\textheight
\epsffile{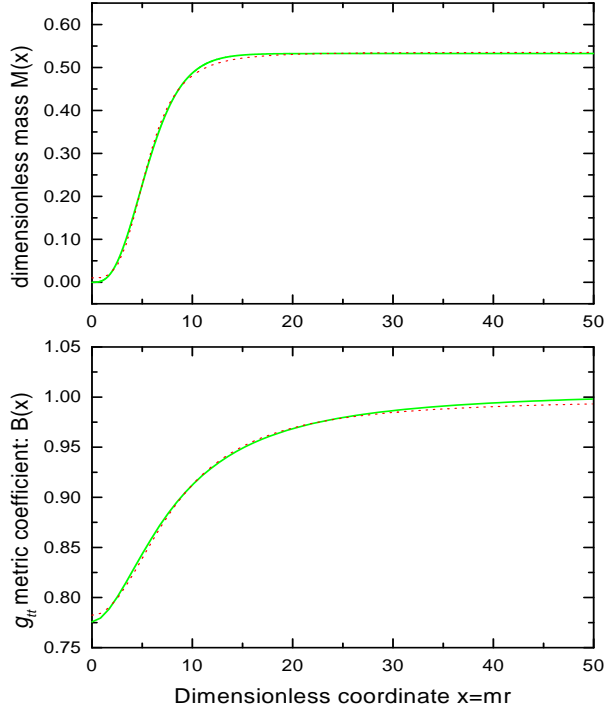} \caption{Logistic function fits to both,
B(x) and the mass. The fit is actually superposed to the numerical
curves in both cases. } \label{fig:fitfig}
\end{figure}

To simplify the following calculations, we adopt a simple Logistic
function to fit both $M(x)$ and $B(x)$ for $x<50$,
\be\label{eq:sys2} M(x)[{\rm
or}~B(x)]=\frac{A_1-A_2}{1+(\frac{x}{x_0})^p}+A_2. \ee For the
mass $M(x)$, the coefficients are $A_1=0.01039$, $A_2=0.53533$,
$x_0=5.49809$ and $p=3.6201$; while for B(x) the coefficients are
$A_1=0.78236$, $A_2=0.99793$, $x_0=8.17558$ and $p=2.09499$. Both
fits are excellent ($\chi^2 < 10^{-5}$) and valid throughout all
space-time (see Fig~\ref{fig:fitfig}); they much improve those
presented in Ref. \cite{Torres:2000dw}. For $x>50$, $M(x)$ and
$B(x)$ are almost exactly the same as those obtained in the
Schwarzschild black hole case, i.e. $M(x)=0.53256$ and
$B(x)=1-2M(x)/x$.

\subsection{Orbits}

The Lagrangian of the geodesics in the space-time with a line
element shown by Eq.~(\ref{eq:sys1}) can be written as (cf.
\cite{chand83}, pages 342-347 --- \cite{msw73}, pages 897-902):
\be\label{eq:sys3} 2{\mathcal
L}=B(r)\dot{t}^2-A(r)\dot{r}^2-r^2(\dot{\theta}^2+
\sin^2\theta\dot{\phi}^2), \ee where the dot denotes
differentiation with respect to an affine parameter $\lambda$
along the geodesics. Together with the assumption of stationary
and axisymmetric space-time, the Euler-Lagrange equations give the
simplified photon momentum equations as (see analogous derivations
of the momentum equations for black hole metric in Ref.
\cite{chand83,msw73}, pages as above):
\begin{eqnarray}
\label{eq:sys4}
\frac{dt}{d\lambda}&=&\frac{r^2}{B(r)}\frac{1}{\sqrt{R}}, \label{eq:sys4a}\\
\frac{dr}{d\lambda}&=&\pm \frac{\sqrt{R}}{r^2},\label{eq:sys4b}\\
\frac{d\theta}{d\lambda}&=&\pm \frac{\sqrt{\Theta}}{r^2},\label{eq:sys4c}\\
\frac{d\phi}{d\lambda}&=&\frac{\xi}{r^2\sin^2\theta},\label{eq:sys4d}
\end{eqnarray}
where the functions $\Theta$ and $R$ are respectively given by
\begin{eqnarray}
\Theta&=&\eta-\xi^2 \cot^2 \theta, \label{eq:sys4e}\\
R&=&\frac{r^4}{A(r)B(r)}-\frac{r^2}{A(r)}(\eta+\xi^2),\label{eq:sys4f}
\end{eqnarray}
and where $\eta={\cal K}/E^2$ and $\xi=L/E$ are constants of
motion related to the angular momentum $L$, the energy at infinity
$E$, and the Carter's constant ${\cal
K}=p^2_{\theta}+L^2\cot^2\theta$. $p_{\theta}$ is the
$\theta$-component of photon momentum (cf. the photon momentum
equations in the space-time of a Kerr black hole
\cite{chand83,msw73}, pages as above).

The path of a particular photon is completely described by the
motion constants $\eta$ and $\xi$, which are in fact related to
the polar and azimuthal angles $\alpha$ and $\beta$ on the sky as
seen by a distant,  locally non-rotating, observer who receives
the photon. The motion constants $\eta$ and $\xi$ are then given
by \footnote{See the derivation in Ref.~\cite{karas92,luyu01} for
the metric of Kerr black hole. The motion constants could be
obtained by replacing the Kerr metric in Karas et al.
\cite{karas92} with the much simple boson star metric. See Fig. 1a
in Ref.~\cite{karas92} for a geometric interpretation of the
angles involved.}
\begin{eqnarray}
\label{eq:sys5}
\xi&=&\left( \frac{r\sin\theta\sin\alpha\sin\beta}{\sqrt{B(r)}}
\right)_{\theta=\theta_{\rm o}},\label{eqs5a}\\
\eta&=&\frac{r^2}{B(r)}\sin^2\alpha\left(1-\sin^2\theta\sin^2\beta\right)_{\theta=\theta_{\rm o}}\label{eqs5b}.
\end{eqnarray}

The equation governing the projection of a photon orbit in ($r$,
$\theta$) plane is \be\label{eq:sys6} \int_{r_i}^{r}
\frac{dr}{\sqrt{R}}=\pm \int_{\theta_i}^{\theta}
\frac{d\theta}{\sqrt{\Theta}}, \ee where $r_i$ and $\theta_i$ are
the initial values of $r$ and $\theta$. The $\theta$-integration,
i.e. the right hand of Eq. (\ref{eq:sys6}), gives
\be\label{eq:sys7} \int
\frac{d\theta}{\sqrt{\Theta}}=\frac{1}{\sqrt{\eta+\xi^2}}
\cos^{-1} \left(\sqrt{\frac{\eta+\xi^2}{\eta}}\cos\theta\right).
\ee The integral for $t$ and $\phi$ can be written as
\begin{eqnarray}
\label{eq:sys8}
t&=&\int \frac{r^2dr}{B(r)\sqrt{R}},\label{eq:sys8a}\\
\phi&=&\int \frac{\xi}{\sin^2 \theta}\frac{d\theta}{\sqrt{\Theta}}
    =\tan^{-1} \left[\sqrt{\frac{1+K}{1-K}} \tan\left(\frac{\theta^{'}}{2}
       \right)\right]
      +\tan^{-1} \left[\sqrt{\frac{1-K}{1+K}}\tan\left(\frac{\theta^{'}}{2}
       \right)\right],\label{eqs:sys8b}
\end{eqnarray}
where $K=\sqrt{\frac{\eta}{\eta+\xi^2}}$, and
$\theta^{'}=\cos^{-1}(\cos\theta/K)$.\\

To enable the integration of a given photon away from the observer
to the disc surface, it is necessary to determine the initial sign
of $dr/d\lambda$ and $d\theta/d\lambda$ which are in fact given
by (see Fig. 1a in Ref.~\cite{karas92}):
\begin{eqnarray}
\label{eq:sys9}
\frac{dr}{d\lambda}&>&0 \ \ \ \  {\rm for}\ \ \ 0\le \alpha < \pi/2,\label{eq:sys9a}\\
\frac{dr}{d\lambda}&<&0 \ \ \ \  {\rm for}\ \ \ \pi/2\le \alpha < \pi,\label{eq:sys9b}\\
\frac{d\theta}{d\lambda}&>&0 \ \ \ \  {\rm for} \ \ \ |\beta| < \pi/2,\label{eq:sys9c}\\
\frac{d\theta}{d\lambda}&<&0 \ \ \ \  {\rm for} \ \ \ |\beta| > \pi/2.\label{eq:sys9d}
\end{eqnarray}
The sign of $dr/d\lambda$ (or $d\theta/d\lambda$) changes at the
turning point where $R=0$ (or $\Theta=0$). \\

Armed with Eqs. (\ref{eq:sys4}) to (\ref{eq:sys9d}), we can trace
the trajectories of photons that propagate from the observer at
($r_\rmo$, $\theta_\rmo$, $\phi_\rmo$) back to the initial
emitting position ($r_{\rm e}$, $\pi/2$, $\phi_{\rm e}$) in the
local rest frame of matter moving in the plane of the disc.

\subsection{Disc}\label{sec:assumcalc}

\begin{figure}[t]
\begin{center}
\epsfxsize=0.5\textwidth \epsfysize=0.35\textheight
\epsffile{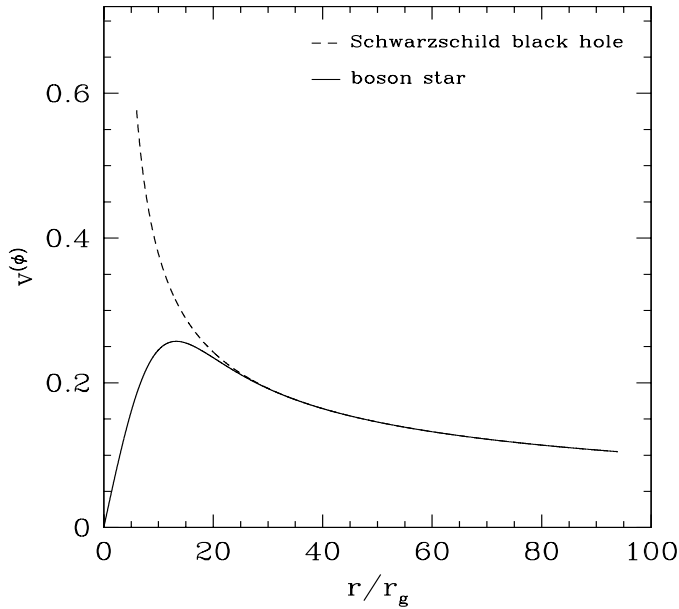} \caption{The rotating velocity of the disc
material in the view of a static observer as a function a radius
(in unit of the gravitation radius $r_{\rm g}=GM/c^2$. The solid
line represents a boson star and the dashed line represents a
Schwarzschild black hole.} \label{fig:figvel}
\end{center}
\end{figure}

In analogy to the standard geometrically thin and optically thick
disc around a black hole, such simplified accretion disc now
surrounding a massive boson star was first discussed in Ref.
\cite{Torres:2002td}. It is assumed that the disc material is in
perfect Keplerian motion which gives the following components for
the 4-velocity (analogous to the derivation for equatorial disk
around Kerr black holes in Ref.~\cite{BARDEEN}):
\begin{eqnarray}
\label{eq:sys10}
u^t&=&\sqrt{\frac{2}{2B(r)-rB^{'}(r)}},\label{eq:sys10a}\\
u^{\phi} &=& r^{-\frac{1}{2}}\sqrt{\frac{B^{'}(r)}{2B(r)-rB^{'}(r)}},\label{eq:sys10b}\\
u^{r}&=& 0,\label{eq:sys10c}\\
u^{\theta}&=&0,\label{eq:sys10d}
\end{eqnarray}
where $B^{'}(r)=dB(r)/dr$. The rotating velocity of the disc
material measured by a static observer is given by \be
v^{(\phi)}=r^{\frac{1}{2}}\sqrt{\frac{B^{'}(r)}{2B(r)-rB^{'}(r)}}.
\ee
 The coordinate-velocity in Eq. (21) is related to
angular velocity by $u^{\phi}=u^t \Omega$.\footnote{Note that
there is a typing error in the first of the Eqs. (3) of Ref.
\cite{Torres:2002td}: $x$ should be in the denominator.} Note that
in the potential of a non-rotating boson star, there are circular
orbits for every possible value of the radial coordinate,
including those which are inside the boson star
\cite{Torres:2002td}. This is clearly different from the black
hole case, where the innermost stable circular orbit only extends
down to $r=6r_{\rm g}$, in the case of a Schwarzschild black hole.
As seen from Figure~\ref{fig:figvel}, at large radii ($r>25r_g$),
the rotating velocity of the disc surrounding a boson star is the
same of that disc surrounding a Schwarzschild black hole. At the
innermost part of the disc ($r<25r_g$), however, differences are
not negligible. In the case of a boson star, with decreasing
radius, the rotating velocity of the disc increases to a maximum
value and then decreases to zero at the star center. In the case
of a Schwarzschild black hole, the rotating velocity of the disc
increases monotonically until the radius approaches the last
stable orbit ($r=6r_g$). To the view of distant observers, then,
this difference results in less Doppler-shifted photons observed
in the putative case of a boson star-accretion disc (BS-AD), as
compared to the Schwarzschild black hole accretion disc
(SchBH-AD).

\subsection{Numerical implementation}

We shall assume that 1) the disc material is not interacting with
the bosons, 2) the distant observer views the disc at some
inclination angle $\theta_{\rmo}$ (with $\theta_{\rmo}=0^0$
corresponding to a face-on disc). Operationally, we define the
distant observer to be located at $r_{\rmo}=10^4 r_{\rm g}$ and
$\phi_{\rmo}=0^0$. We then integrate the four photon momentum
equations from the observer to the accretion disc using motion
constants $\xi$ and $\eta$. The redshift factor of a photon
propagating from the disc to the observer is given by \be
g=\frac{\nu_{\rm o}}{\nu_{\rm e}}= \frac{{\bf p}_{\rm o}\cdot {\bf
u}_{\rm o}}{{\bf p}_{\rm e}\cdot {\bf u}_{\rm e}}, \ee where
$\nu_{\rm o}$ and $\nu_{\rm e}$ are the frequencies of the photon
received by the distant observer and the photon emitted in the
rest frame of disc material, respectively; ${\bf p}_{\rm o}$ is
the photon momentum received by the distant observer and ${\bf
p}_{\rm e}$ is the photon momentum in the rest frame of the
emitting material. The disc region between $r_{\rm out}$ and
$r_{\rm in}$ is divided into $1000\times 1000$ elements
($r_i\rightarrow r_i+dr$, $\phi_i\rightarrow\phi_i+d\phi$). The
solid angle subtended at the observer's sky by each element is
also calculated by counting the photons impinging onto the element
(we refer the reader to Refs. \cite{luyu01,yulu00} for
further details).\\

Similar to the case of an accretion disc surrounding a black hole,
the disc rotating around a boson star may also be illuminated by
some nearby hard X-ray source due to similar mechanism, for
example, inverse Compton scattering of soft thermal photons by
high energy electrons in the corona above the disc. The
fluorescent Fe K$\alpha$ line photons driven by this illumination
may directly propagate to distant observers without absorption. We
assume, as it is usual, that the X-ray illumination is
axisymmetric and the emissivity of Fe K$\alpha$ line only varies
with radius as a power law $\epsilon(r)\propto r^{-\alpha}$ in the
rest frame of emitting material, where $\alpha$ is the emissivity
index. We already know that the viscosity dissipated power per
unit area approaches a maximum value at the radius of $\sim 5
r_{\rm g}$ and then rapidly decreases as the accreted material
goes into the star center \cite{Torres:2002td}. This is quite
different from the case of black hole accretion disc, where the
radiation from the region within the marginally stable orbit (in
projection) is
insignificant.\\

In order to see the difference between the profiles of the
emission lines coming from a BS-AD and from a SchBH-AD,
respectively, we consider two cases. In the first one, for a
BS-AD, the line is assumed to be emitted from an annulus between
$r_{\rm in}=6r_{\rm g}$ and $r_{\rm out}=30 r_{\rm g}$. In the
second case, the line will be assumed to come from an annulus
between $r_{\rm in}=2r_{\rm g}$ and $r_{\rm out}=30r_{\rm g}$. At
all times, for a SchBH-AD, the line is assumed to be emitted from
annulus between $r_{\rm in}=6r_{\rm g}$ and $r_{\rm out}=30r_{\rm
g}$. Physical considerations can be put forward to justify these
values of inner and outer limits of the disc. In the case of
BS-AD, $r_{\rm in}$ is not chosen to be closer to the center (or
0) because we are adopting a power-law emissivity law which
diverges at the star center. {\it We note, however, that the line
emission within $r \leq 2r_{\rm g}$ would be negligible if the
emissivity law is similar to the viscosity dissipated power as a
function of radius.} We set $r_{\rm in}=6r_g$ in the case of
SchBH-AD since the disc only extends to the marginal stable orbit.
The line emission from the disc material within the marginal
stable orbit is generally assumed negligible, and if this line
flux is even observable, as Reynolds and Begelman \cite{rb97} have
pointed out, it would generally emerge in the extremely red wing
of the line profile, what would make the difference between the
profiles of Fe K$\alpha$ generated from the BS-AD and from the
SchBH-AD that we present in the next Section larger.\\

The observed line flux distribution is $dF(E_{\rm obs})=I(E_{\rm
obs})d\Omega$, where $d\Omega$ is the solid angle subtended by the
disc in the observer's sky and $I(E_{\rm obs})$ is the specific
intensity received by the observer.  The specific intensity
$I(E_{\rm em})$ of the line emitting in the rest frame of disk
material is assumed to be isotropic and monochromatic $I(E_{\rm
em})=\epsilon(r)\delta(E_{\rm em}-E_0)$, where $E_0$ ($=6.4$ keV
for iron K$\alpha$ line) is the energy of the line photons in the
rest frame of the emitting material and $gE_{\rm em}=E_{\rm obs}$.
From the relativistic invariance $I_{\nu}/\nu^3$ (see explanations
in Ref.~\cite{msw73}, pages 897-902), we have $I(E_{\rm
obs})=(\nu_{\rm o}/ \nu_{\rm e})^3 I(E_{\rm em})=g^3 I(E_{\rm
em})=g^4 \epsilon(r)\delta(E_{\rm obs}-gE_0) $, and thus the total
observed flux distribution of the line is given by \be F(E_{\rm
obs})\propto \int_{r_{\rm in}}^{r_{\rm out}}\epsilon(r)
g^4\delta(E_{\rm obs}-gE_0)d\Omega_{\rm obs}, \ \ \ \ (r_{\rm
in}\le r \le r_{\rm out}). \ee This is what is plotted in the
graphs of next Section.

\section{Line Profiles}\label{sec:profs}

Theoretical emission line profiles produced in the inner region of
accretion discs around black holes have been calculated by several
authors, see for instance Refs. \cite{fabian89,laor91,fanton97}.
The analysis of Section~\ref{sec:photon} and
Section~\ref{sec:assumcalc} has shown that there are two main
factors producing differences between the line profiles emitted
from the disc extending down within a boson star or the usually
considered one accreting around a Schwarzschild black hole:

\begin{enumerate}
  \item The gravitational potential of a boson star is less deeper
than that of a Schwarzschild black hole in the innermost region
($r<15r_{\rm g}$).
  \item The rotation velocity of the disc around a boson star is
smaller than that of the disc around a Schwarzschild black at the
inner region with $r<25 r_{\rm g}$
\end{enumerate}

\begin{figure}[t]
\begin{center}
\epsfxsize=0.8\textwidth \epsfysize=0.80\textheight
\epsffile{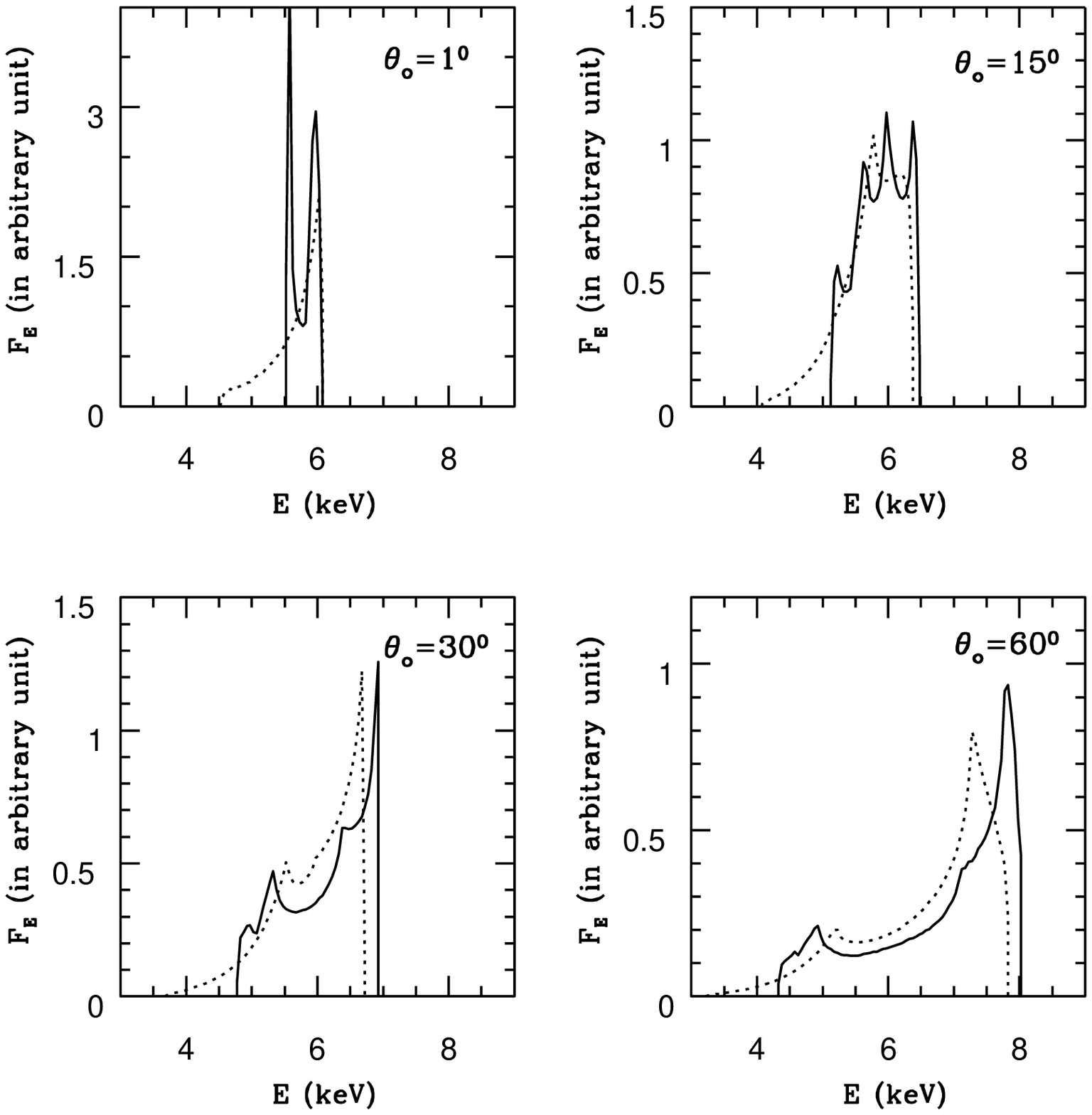} \caption{Line profiles computed using
complete relativistic method as described in the text. The line
emitting region for all the cases is fixed to be the annulus with
$r_{\rm in}=6r_{\rm g}$ and $r_{\rm out}=30r_{\rm g}$. The line
emissivity law is set to be $\epsilon\propto r^{-2}$ and the
inclination angles are labeled in each panel. The solid lines and
dashed lines represent the profiles of lines from boson
star-accretion disc and from Schwarzschild black hole-accretion
disc, respectively. } \label{fig:a2in6}
\end{center}
\end{figure}

\begin{figure}[t]
\begin{center}
\epsfxsize=0.8\textwidth \epsfysize=0.80\textheight
\epsffile{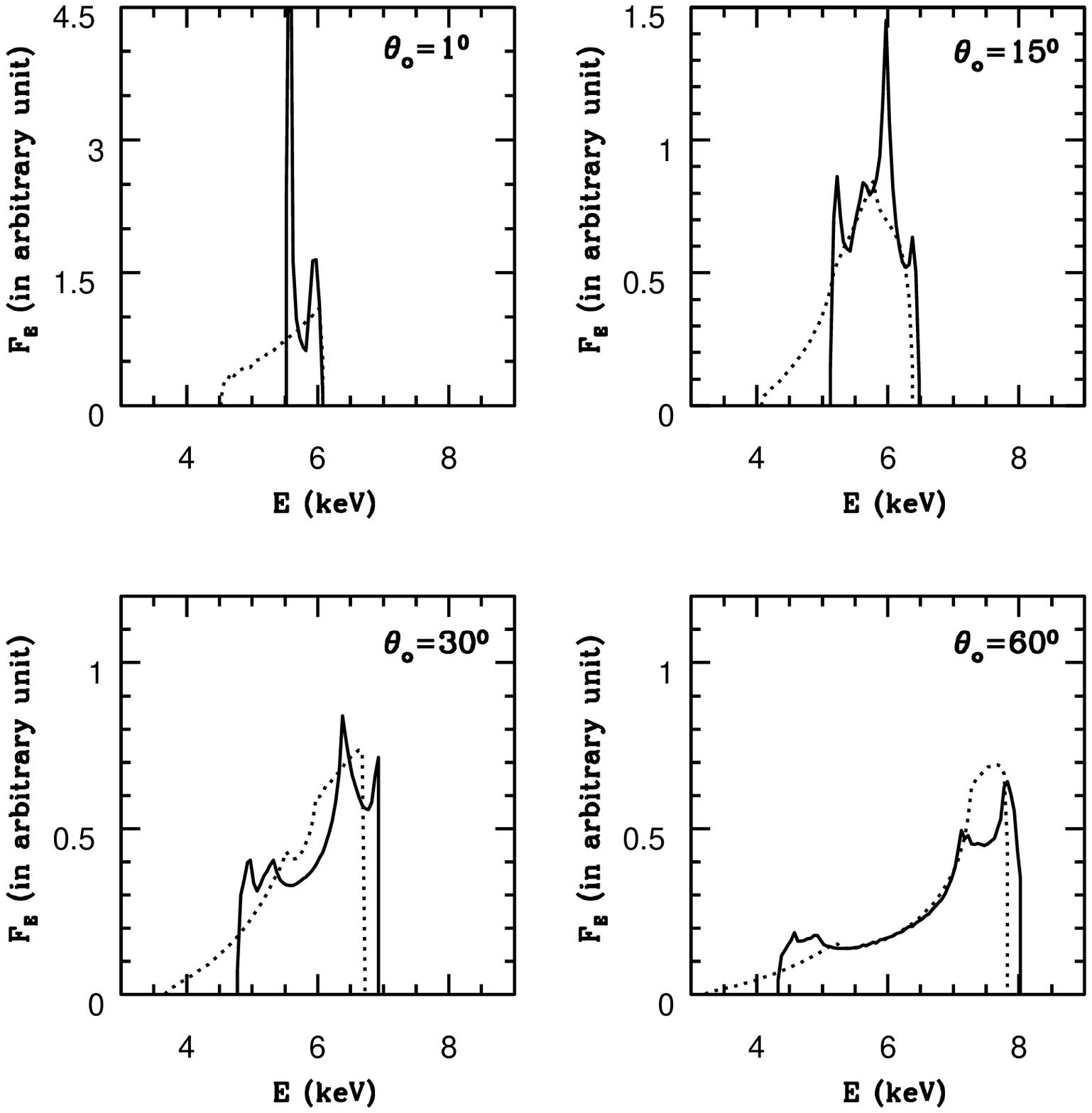} \caption{Same as Fig.~\ref{fig:a2in6}, but
adopt a different emissivity law of $\epsilon \propto r^{-3}$. }
\label{fig:a3in6}
\end{center}
\end{figure}

\begin{figure}[t]
\begin{center}
\epsfxsize=0.8\textwidth \epsfysize=0.80\textheight
\epsffile{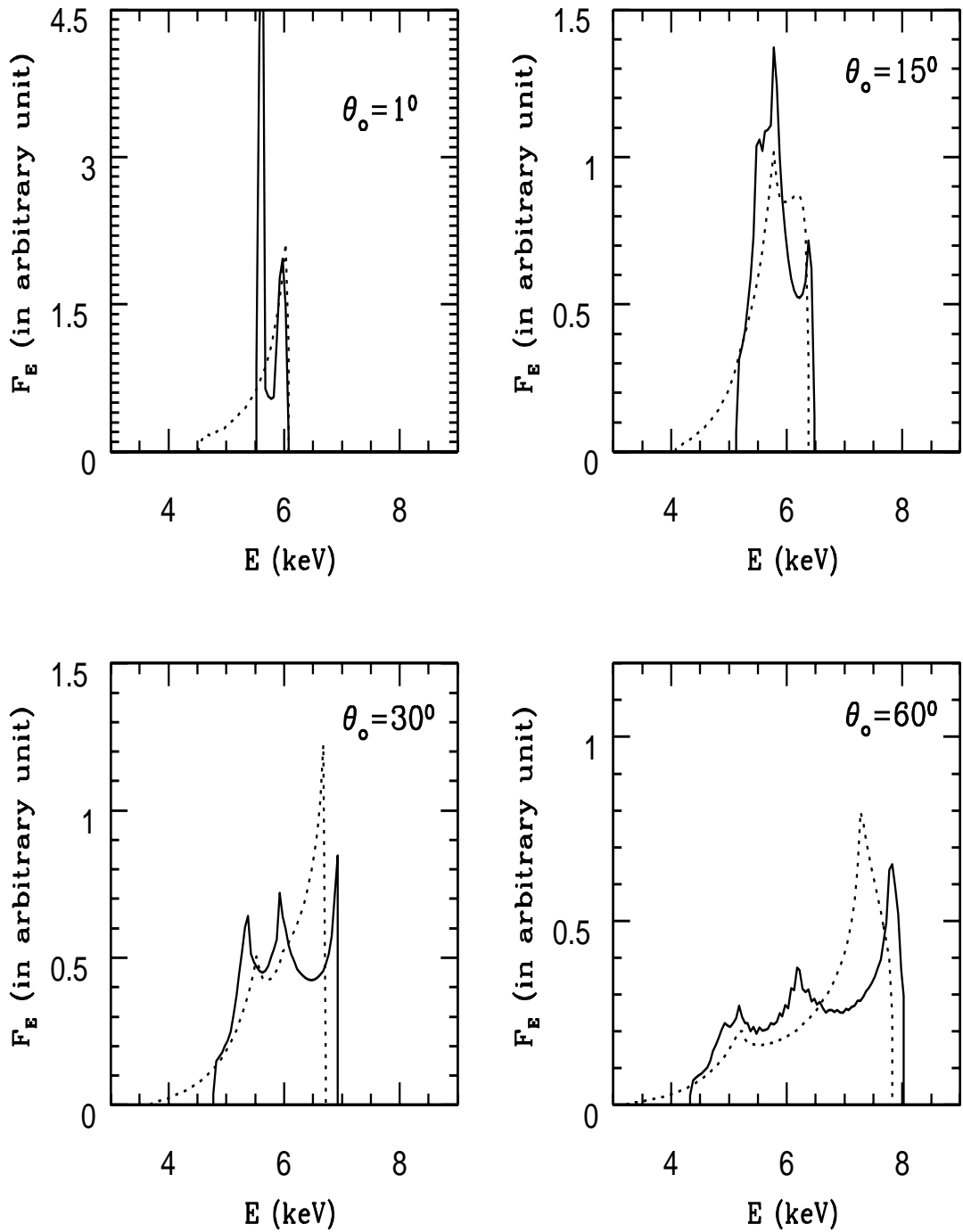} \caption{Same as Fig.~\ref{fig:a2in6}, but
adopt a different inner radius $r_{\rm in}=2r_{\rm g}$ of the line
emitting annulus for the case of the boson star-accretion disc. }
\label{fig:a2in2}
\end{center}
\end{figure}

\begin{figure}[t]
\begin{center}
\epsfxsize=0.8\textwidth \epsfysize=0.80\textheight
\epsffile{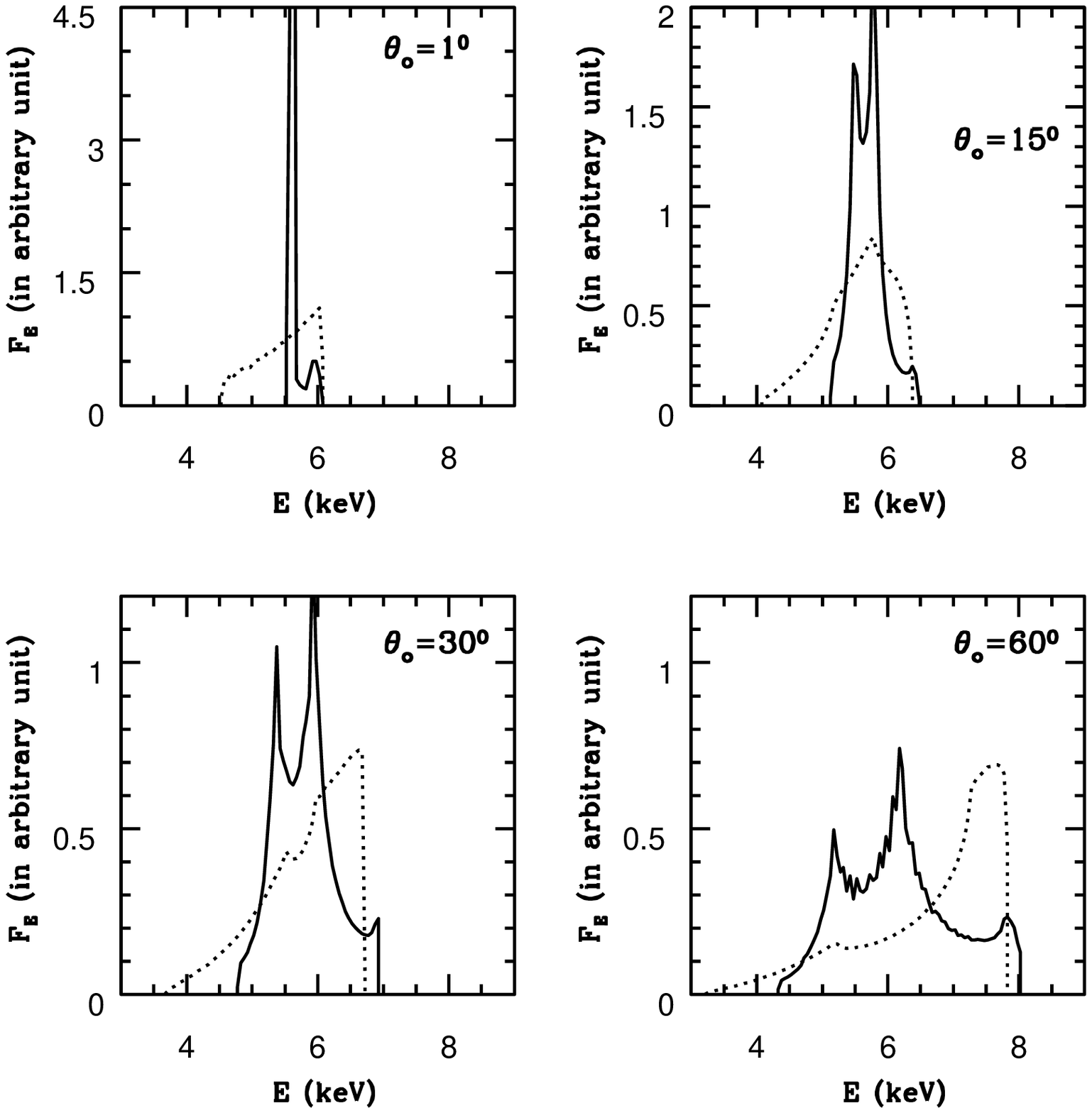} \caption{Same as Fig.~\ref{fig:a2in2}, but
adopt a different emissivity law of $\epsilon \propto r^{-3}$. }
\label{fig:a3in2}
\end{center}
\end{figure}

Figures~\ref{fig:a2in6}-\ref{fig:a3in2}  illustrate the line
profiles in these two different situations using a completely
relativistic calculation with a large range of parameters. Details
of the code used can be found in references \cite{luyu01,yulu00}.
These profiles correspond to the inner disc annulus with $r_{\rm
in}=6r_g$ and $r_{\rm out}=30r_g$ (Fig.~\ref{fig:a2in6} for
$\alpha=2$ and Fig.~\ref{fig:a3in6} for $\alpha=3$) or with
$r_{\rm in}=2r_g$ and $r_{\rm out}=30r_g$ (Fig.~\ref{fig:a2in2}
for $\alpha=2$ and Fig.~\ref{fig:a3in2} for $\alpha=3$) in the
case of a BS-AD (solid lines); or with $r_{\rm in}=6r_g$ and
$r_{\rm out}=30r_g$ in the case of a SchBH-AD
(dotted lines). All the lines are normalized by their line flux.\\

These are the main features of our results:

\begin{itemize}
  \item The lines emitted from the inner region of a BS-AD are
somewhat narrower than those emitted from the inner region of a
SchBH-AD. The red wing of the lines emitted from the inner region
of a BS-AD does not extend to extremely low energy as that from a
SchBH-AD because 1) the gravitational shifting of line photons (a
factor of $\sqrt{B(r)}$) emitted from the inner region of BS-AD is
much less than that from the inner region of SchBH-AD (see
Fig.~\ref{fig:gttfig} and 2) the Doppler redshifts of line
photons in the case of BS-AD is also less than that in the case of
SchBH-AD, since the rotating velocity of the inner disc is much
smaller in the former case (see also Fig.~\ref{fig:figvel}). The
high energy blueward extent of the lines emitted from the inner
region of BS-AD is determined by the inclination angle of the
disc, which is similar to the case of SchBH-AD. However, for the
same inclination angle, in the case of BS-AD, the blue horn
extends to a larger energy than that in the case of SchBH-AD,
since the former case suffers less gravitational redshifts.

  \item Generally, the line profiles emitted from BS-AD appears to have
more peaks than that emitted from SchBH-AD (see
Figs.~\ref{fig:a2in6}-\ref{fig:a3in2}). For very small inclination
angles (e.g. $1^0$, see upper left panel in
Figs.~\ref{fig:a2in6}-\ref{fig:a3in2}), the line emitted from the
inner region of BS-AD is double peaked, and the low energy (more
intense) peak is due to the emission from the inner edge where the
gravitational shifting is the dominant factor (the transverse
Doppler shifting is not large at the inner edge). The line from
the inner region of SchBH-AD, on the contrary, tends to a single
broad peak (and less intense) with skewed red wings due to much
deeper gravitational potential and larger rotating velocity of the
disc material near the marginal stable orbit.

  \item In discs with large inclination angles (e.g. $>15^0$), the
line profiles emitted from the inner region of BS-AD may have more
peaks. A typical case with inclination angle $\theta_{\rm o}=30^0$
is illustrated in the left bottom panel in Fig.~\ref{fig:a2in6}:
the profile for the inner region of a BS-AD has totally four local
maxima compared to the only two local maxima for the case of
SchBH-AD. From right to left, the first and the third maximum are
due to the line emission in the outer part of the annulus [with
$r(v\sim v_{\rm max})$]. There are small displacements between
these two maxima and the corresponding two maxima in the case of
SchBH-AD. The second and the fourth maximum are due to the line
emission from the inner part of the annulus [with $r(v\simlt
v_{\rm max})$]. There are no such two corresponding peaks in the
case of SchBH-AD because combined effects of the much deeper
gravitational potential of the black hole and the larger disc
rotation velocity shifts the line photons to lower or higher
energy and smooth the line profile to the skewed one.
\end{itemize}


\section{Concluding remarks}

Observations of the central regions of active galaxies already
exist. For some of them, the Iron K$\alpha$ profile could be
measured and interesting conclusions can be already drawn. For
instance, given the observations of MCG-6-30-15, in which the red
wing of the Fe K$\alpha$ extends down to energy of 2-3 keV
\cite{wilm01}, the computations of this paper have clearly shown
that the central object in this galaxy cannot be a non-rotating
massive boson star. It would be of interest, though, to study more
complicated cases of accretion discs, particularly those around
rotating boson stars. The observational data will be very much
improved with forthcoming technologies. The NASA Constellation-X
\cite{consx} mission, to be launched in 2008, is optimized to
study the iron K$\alpha$ line features and will determine the
black hole mass and spin for a large number of systems. A
detection of several peaks at lower energies, or of intense double
peaks in faced on galaxies could be an impressive indication of a
boson star nucleus. More importantly, this paper gives a new way
to examine an interesting theoretical construct with real
astrophysical data, helping to elevate the boson star concept to a
falsifiable level.

\section*{Acknowledgments}

YL thanks the hospitality of the Astrophysical Sciences Department
of Princeton University. DFT acknowledges support from Fundaci\'on
Antorchas, CONICET, and the Physics Department of Princeton
University.



\end{document}